# Atomic-scale electron-beam sculpting of defect-free graphene nanostructures

*Bo Song, Grégory F. Schneider, Qiang Xu, Grégory Pandraud, Cees Dekker, Henny Zandbergen\**

Kavli Institute of Nanoscience, Delft University of Technology, Lorentzweg 1, 2628 CJ Delft, The Netherlands

**\*Corresponding author,** h.w.zandbergen@tudelft.nl

**ABSTRACT.** In order to harvest the many promising properties of graphene in (electronic) applications, a technique is required to cut, shape or sculpt the material on a nanoscale without damage to its atomic structure, as this drastically influences the electronic properties of the nanostructure. Here, we reveal a temperature-dependent self-repair mechanism allowing damage-free atomic-scale sculpting of graphene using a focused electron beam. We demonstrate that by sculpting at temperatures above 600 °C, an intrinsic self-repair mechanism keeps the graphene single-crystalline during cutting, even thought the electron beam induces considerable damage. Self-repair is mediated by mobile carbon ad-atoms constantly repairing the defects caused by the electron beam. Our technique allows reproducible fabrication and simultaneous imaging of single-crystalline free-standing nanoribbons, nanotubes, nanopores, and single carbon chains.

**MANUSCRIPT TEXT.**

Graphene, a one-atom thin sheet of carbon atoms, is the building block of fullerenes, carbon nanotubes, nanoribbons and graphite. First isolated by Geim and Novoselov in 2004,[1-3] graphene has received tremendous scientific attention due to its unique electronic properties.[4] Graphene also features edge dynamics[5] and mechanical properties,[6] opening up even more opportunities such as its use to sequence genomic DNA using nanopores[7-9] and nanogaps[10]. Having a zero band gap, graphene cannot be used directly in electronic applications such as field-effect transistors,[11] but theoretical studies showed that a band gap can be opened by confining the graphene to for instance a ribbon,[12,13] whereby the crystallographic edges and width of the ribbon determines its electronic properties. Current techniques to fabricate a ribbon (or any given shape),[14-17] however, lack the required sub-nanometer precision for obtaining atomically sharp and controlled regular edges, with an appropriate crystal orientation.

The required atomic precision could in principle be obtained with a focused electron beam of a transmission electron microscope. It is long known that such an electron beam offers the required precision, but it is not used because it generates undesired defects and artefacts, such as fast amorphisation of the graphene crystal structure and carbon deposition.[18-19] Generally 80 keV electrons are used to prevent electron beam damage.[5] However, sculpting of graphene can only be done efficiently by knocking out carbon atoms from the graphene lattice, requiring an electron energy above 140 keV.[20] But electrons with such energy are known to also amorphisize graphene, yielding a poorly crystalline lattice. As we will show in this paper, this amorphisation surprisingly does not occur at high temperature, which is a key element to open a route to atomically précis sculpting without inducing artefacts or non-crystallinity.

To perform the experiments, we used a FEI Titan transmission electron microscope operated at 80 and 300 keV, with electron doses of $10^5$ and $10^7$ electrons/nm$^2$s for respectively imaging and sculpting. To heat the graphene, we used a home-made MEMS-based sample holder (details in Supporting Information, SI),[21] on which graphene was deposited using wedging transfer (SI Figure S1).[22]

Temperature has a remarkable effect on the changes induced by 300 keV electrons (Figure 1). At room temperature (RT), a rapid amorphisation occurs (Figure 1a), which prevents detailed high-resolution electron microscopy (HREM). Thus, sculpting of fully crystalline graphene nanostructures cannot be done with 300 keV electrons at RT. In addition to amorphisation, some residual C deposition on the graphene occurs even if the sample is pre-heated or pre-annealed (SI Figure S2). At temperatures of 200 °C, the C deposition is slow enough to study the temperature-dependent atom rearrangement in grahene. We observe that electron beam irradiation leads to amorphisation with only short range order (Figure S3). At 500 °C, the electron beam results in the formation of polycrystalline monolayers (Figure 1c and Movie S1). The single crystalline graphene transforms into polycrystalline graphene with clear straight but short grain boundaries. At 700 °C, remarkably, graphene conserves its full crystallinity even under a very intense electron beam (Figure 1d). Fourier transforms of the images in Figure 1 confirm this trend from fully amorphous RT to fully crystalline at 700 °C (Figure S3). Areas that are made amorphous at room temperature are found to convert into (poly)crystalline by the combination of high temperature and electron beam irradiation (SI Figure S4). Since the C-atom knockout probability does not depend on temperature, the absence of amorphisation at higher temperatures points to a self-repair process.

We identified that in this self-repair mechanism the vacancies that are created by the knock-out, are quickly reoccupied by C ad-atoms which are present on the surface originating from existing C-rich contaminations or from the knockout in the area exposed to the electron beam. C ad-atoms diffuse over the surface of graphene (even at RT),[23] and are trapped by defects in the graphene as shown by red arrows in Figure 1c and Figure 1d.

Self-repair in graphene is best at temperatures above 600 °C, as illustrated in Figure 2. A highly defective area including lattice rotations is created with a very intense 3nm wide electron beam in a graphene monolayer (blue dashed circle in Figure 2a is at 80 % of maximum beam intensity). The lattice spontaneously heals over a time-span of 20 seconds, wherein the graphene recovers its single-crystallinity (Movie S2).

After a vacancy is created in a defect-free lattice by knockout of an atom by the e-beam, various processes can occur, each with their own temperature-dependent probabilities. First, a carbon ad-atom can fill the vacancy site. Second, the vacancy can diffuse into the hole created during the sculpting. Both these processes result in self-repair. Third, the graphene lattice around the vacancy can reconstruct forming new C-C bonds, within one graphene layer,[23] between graphene layers,[24] or with ad-atoms[23]. Fourth, vacancies can congregate,[25] inducing more extended lattice reconstructions.[23] Lattice reconstructions will make self-repair less likely to occur, because an energy barrier has to be overcome to recreate the defect-free lattice. A sequence of lattice reconstructions generated by repeated knockouts of C atoms, can result in amorphisation. The probabilities and times needed for lattice reconstruction and vacancy annihilation will determine whether self-repair or amorphisation dominates. Our results show that amorphisation dominates below 400 °C and that self-repair dominates above 600 °C. Next, we applied graphene self-repair to sculpt shapes that are of interest for fundamental studies and applications.

First, we fabricated graphene nanopores (formation of nanopores is very fast, about 10 seconds). As mentioned in Figure 1, specimen temperature determines whether the surrounding of the formed nanopores is amorphous or crystalline. Pores as small as ten hexagons could be made (e.g., 7 Å, Figure 3a). Depending on the intensity distribution in the electron beam, a straight cylindrical hole (Figure S5) or a very shallow hole surrounded by a number of terraces can be sculpted (Figure 3b). As can be seen in Figure 3b, most edges in the multilayer graphene show a dark rim indicating multiple C atoms in projection, a consequence of curving, connected adjacent graphene layers,[26] or back folding.[27,28] The images of the actual edges of the nanopores in single layers (cf. Figure 1c, 1d, and 3b) do not show, however, any enhanced contrast indicative of such curvature. The terraced layout of the nanopore in Figure 3b, combines the advantage of a monolayer nanopore with the increased stability of the multilayer support.

Second, we sculpted graphene nanobridges. In this case, we start by making two elongated holes. Once the bridge was formed, it can be further modified: depending on the number of graphene layers that one starts sculpting into, one typically forms cylindrical nanotubes, which could be triple, double or single walled (Figure S6, Movie S3), or one can create a flat carbon nanotube as shown in Figure 3c (Movie S4). With continued electron beam irradiation, such bridges change in shape and width.

Third, we sculpted graphene nanoribbons (at lower temperatures, nanoribbon are polycrystalline; SI Figure S7 and Movie S5). However, due to self-repair, nanoribbons are crystalline, straight, and adopt armchair-type edges (Figure 3d, Movie S6). Image simulations (see inset in Figure 3d) show that the dark spots on the edge of the ribbon are mostly due to C ad-atoms. We observed that at 700 °C and under 300 keV electrons, armchair edges are more stable than zigzag edges, in agreement with quantum mechanical calculations.[29] At room temperature and under 80 keV electrons, similarly to Girit et al, we also observed long zigzag edges.[5]

Fourth, we made carbon chains. We started from a nano-ribbon (Movie S7). We first formed a double carbon chain (Figure 3e), which we reconfigured later into a single chain (Figure 3f). Single chains were observed to change their anchor sites frequently, while double chains are in general shorter than the single carbon chains (e.g., 1 nm *vs* up to 3 nm in length, respectively). Based on many movies we suggest that the two edges of the bridge remain in the case of formation of a double chain, whereas the atoms in the middle of the bridge disappear, indicating that very narrow ribbons (i.e., 2 to 3 hexagons wide) are less stable than two chains.[30]

In summary, with a high-voltage e-beam at temperature above 600 °C, graphene can be sculpted into any given shape at single-hexagon resolution, while remaining structure entirely crystalline. Key for achieving this is to evoke the self-repair properties of graphene at high temperature, which counteract the unwanted amorphisation induced by the electron beam. Using this approach we sculpted nanopores, nanotubes, nanoribbons, and single carbon chains. This electron microscopy approach to nanofabrication allows modification on the atomic scale as well as direct visual inspection.

**ACKNOWLEDGMENT.** We thank Vasili Svechnikov, Vlad Karas, Mengyue Wu and Victor Calado for technical assistance. This work was supported by Nano-IMaging under Industrial Conditions (NIMIC) and funding from the European Union's Seventh Framework Programme (FP7/2007-2013) under grant agreement n° 201418 (READNA).

**FIGURE CAPTIONS.**

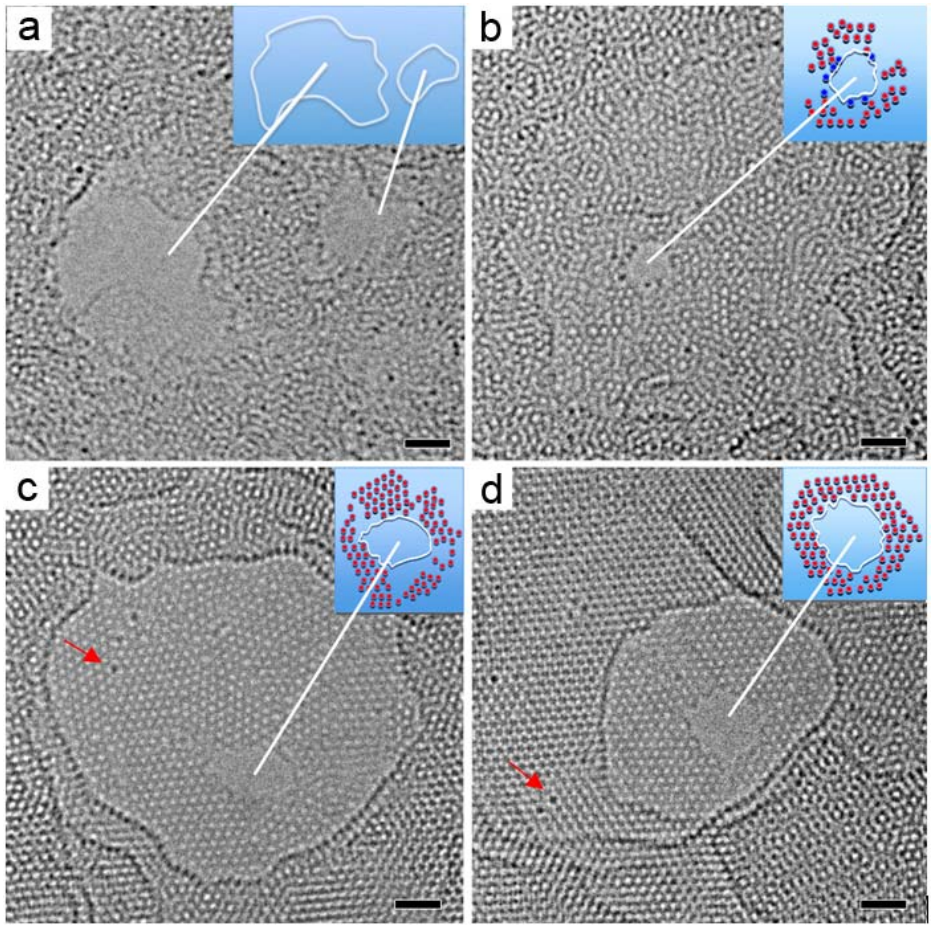

**Figure 1.** Influence of temperature on the sculpting of few-layer graphene by an electron beam. (a), at temperature (RT). (b), at 200 °C. (c), at 500 °C. (d), at 700 °C. At RT two holes are formed (of which only the left one was intended). The whole area surrounding the holes has become amorphous. In the experiments resulting in images (b)-(d), we first removed several graphene layers, a procedure that is very reproducible at 500 °C and 700 °C, but hard to control and verify at 200 °C. Local irradiation was continued until a small hole was formed. The remaining monolayer is almost amorphous at 200 °C, polycrystalline at 500 °C, and single crystalline at 700 °C. Red arrows indicate some of the C ad-atoms trapped at defects. The insets in (b)-(d) show the positions of the identifiable hexagons (red dots) and the estimated position of the edge (white line). The blue dots in the inset at 200 °C are ad-atoms. Scale bars 1 nm.

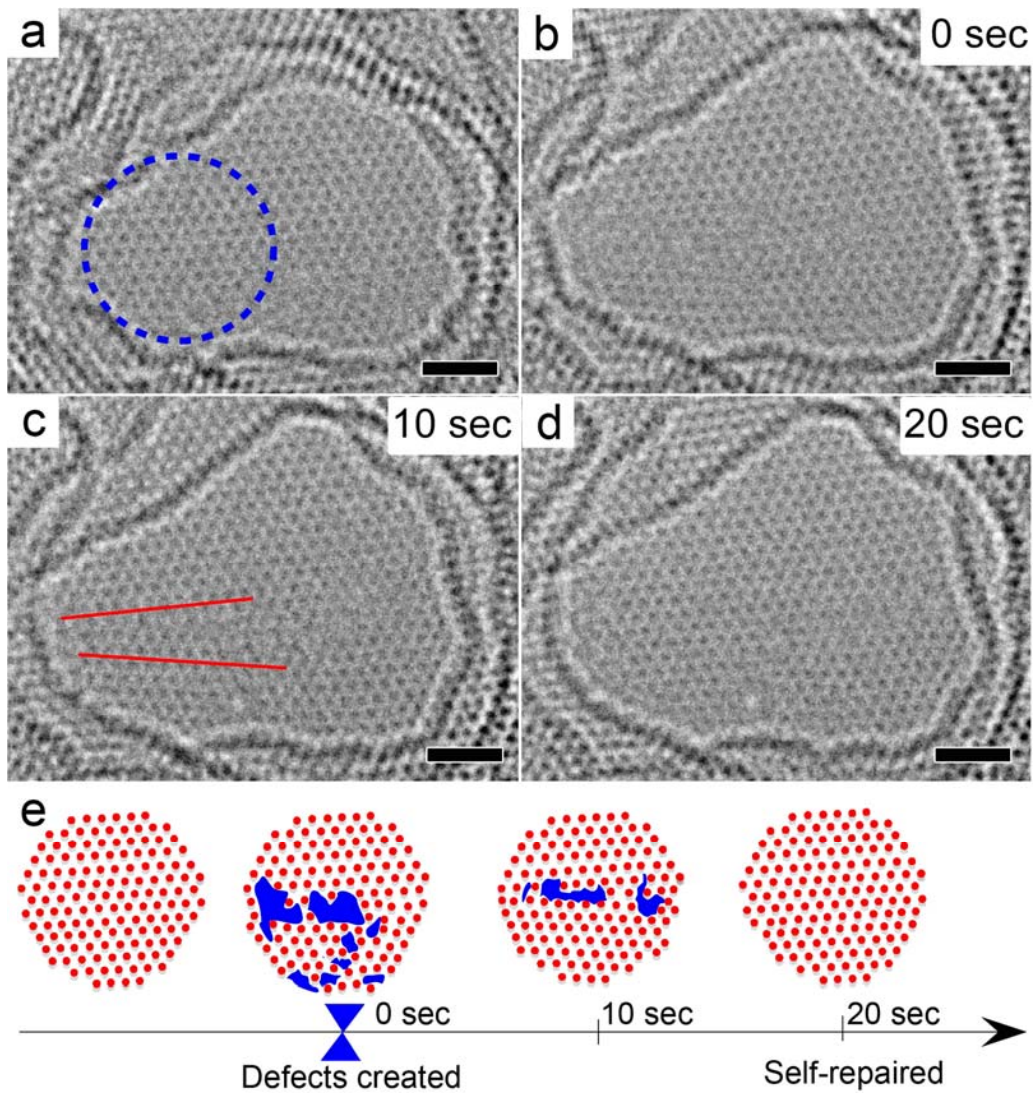

**Figure 2.** HREM images illustrating the self-repair occurring in a graphene monolayer at 600 °C. (a) Crystalline graphene lattice with a monolayer in the central part of the image before irradiation with an intense e-beam. (b), A highly defective area is created by a very intense e-beam at a crystalline area indicated by the blue circle in a). Due to the intense e-beam, the regular lattice has partly amorphisized and part of lattice has rotated. (c)-(d) Over a time-span of ~20 seconds, the graphene subsequently recovered its single-crystallinity. (e) Illustration of the self-repair process. The centers of the C 6-rings (visible in the HREM images as dark dots), are represented as red dots. The blue area indicates where these black dots are absent. Scale bars 1 nm.

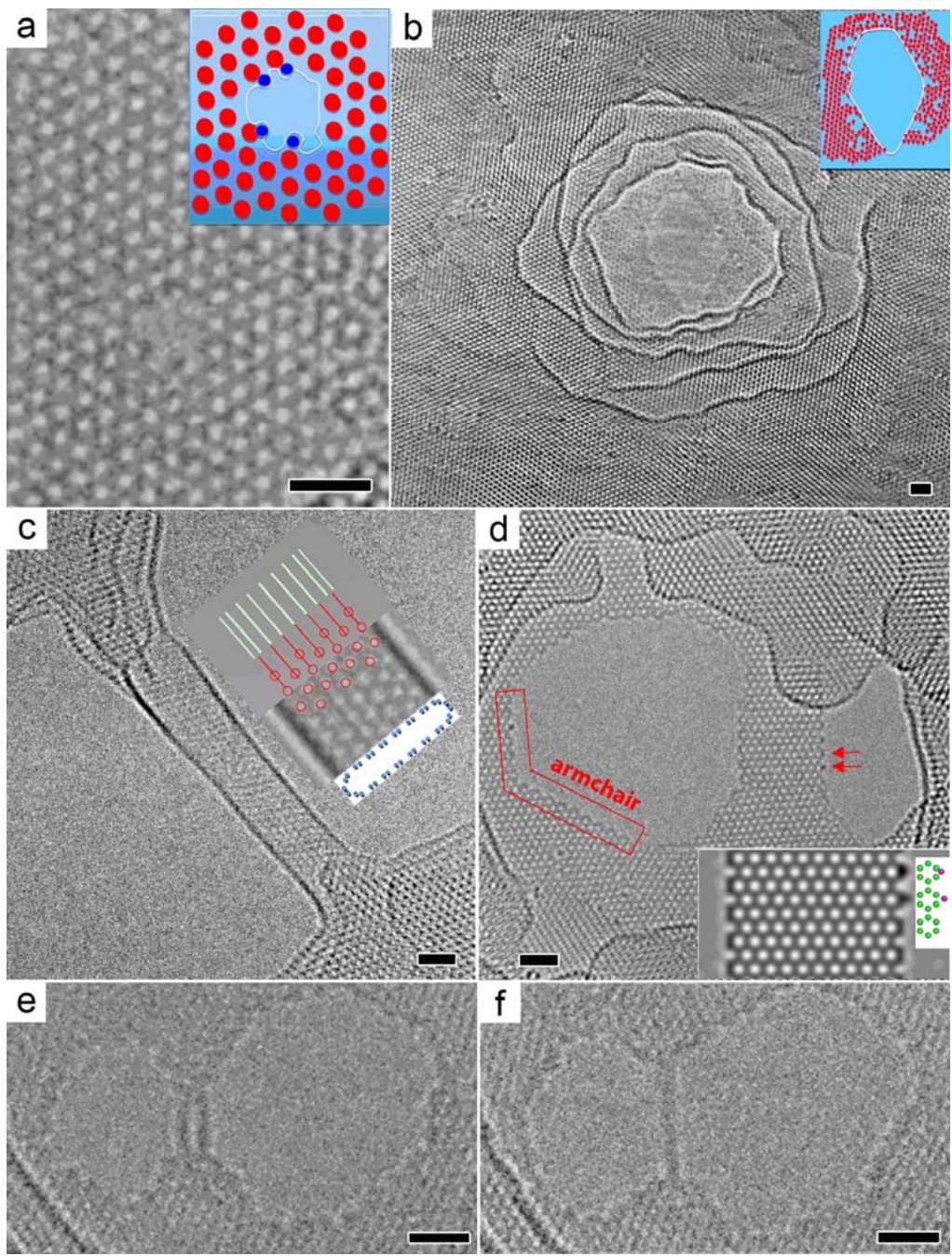

**Figure 3.** Variety of crystalline carbon nanostructures made with 300 kV electrons at 600-700 °C. (a) A 7 Å nanopore was made on a monolayer graphene at 700 °C. (b) A shallow nanopore with terraces in ~10 layers graphene. Insets in a-b represent overlays with red dots depicting hexagons and the white line depicting the estimated nanopore edge. (c) Flat nanotube made from few layers graphene. The inset gives the inferred tube shape with the HREM image averaged by translation, and the observed (red) and estimated (green) hexagon positions for a round tube. Blue dots in the side view represent individual

carbon atoms. (d) Nanoribbon made from single layer graphene, where the monolayer was first made from few layers graphene. Note that the edges of the nanoribbon are armchair, as are most of the edges of the two holes. Inset shows the image simulation of two carbon ad-atoms attached to the graphene armchair edges and pointed out with red arrows. (e)-(f) Two images from Movie S7, showing a double carbon chain and a single carbon chain (formed from the double chain), respectively. Scale bars 1 nm.

**REFERENCES.**

**Supporting Information**

**Content**

- Heating holder that was used

- Sample preparation

- Parameters for TEM sculpting and imaging

- Contamination

- Fourier transforms of images in Figure 1

- Recrystallisation at T>500 °C after amorphisation at room temperature

- Straight cylindrical nanopores

- Carbon nanotubes

- Polycrystalline graphene nanoribbons at 500 ℃

- Prospective: More efficient sculpting of graphene by better hardware

- List of movies supplied

**Heating holder with MEMS heater used for in-situ experiments**

For the in-situ experiments, a SiN membrane was used with an embedded, coiled Pt wire, see also Fig.S1 [S1]. In the SiN membrane, a 5 μm diameter hole was made with a focussed ion beam in between the windings of the Pt wire to allow substrate-free TEM imaging of the graphene. The very low heat capacity of the heater results in low thermal drift, which enables stable high-resolution electron microscopy at elevated temperature. We were able to take images with exposure times up to 10 seconds without loss of atomic resolution due to specimen drift.

**Sample preparation**

Graphene flakes were prepared by exfoliation of natural graphite (NGS graphite) on a 285 nm thermally grown $SiO_2$/Si wafer (Nova electronic materials). Graphene flakes of interest were selected using optical interference microscopy (Fig.S1a) [S2]. Such a graphene flake was then transferred on top of the hole in SiN membrane (see previous section) using the wedging transfer technique [S3] (Fig. 1b). For this, first, a hydrophobic polymer (cellulose acetate butyrate dissolved in ethyl acetate, 30 mg/mL) was deposited on the Si-$SiO_2$-graphene ensemble. Subsequently this polymer coating with the graphene molded into it was wedged at the air/water interface, yielding a floating film at the air/water interface. Next, the MEMS microheater was positioned under the film, and brought into contact with graphene by lowering the level of the water. Submicron positioning was carried using micromanipulators. Finally the polymer was dissolved in pure ethyl acetate.

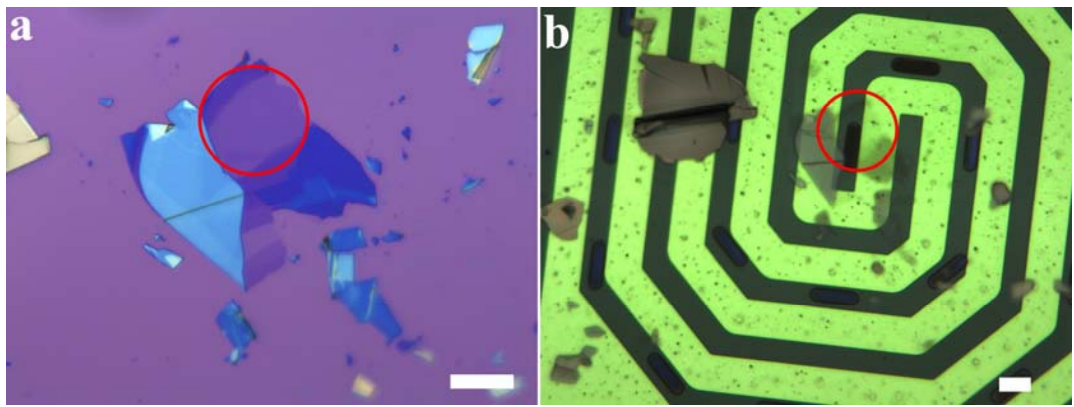

**Figure S1.** Transfer of graphene to a MEMS heater. (a) Optical image of graphene flakes on a $SiO_2$-coated Si wafer. (b) Optical image of the same graphene flake (marked by red circle) after transfer on top of the hole on the Pt microheater. Scale bars: 10 μm.

**Parameters for TEM sculpting and imaging**

The TEM imaging was performed in a cubed FEI Titan microscope with a post-specimen corrector. The spherical aberration is always corrected to below 1 micron. The entire experiment was conducted in the high vacuum environment ($10^{-8}$~$10^{-7}$ torr) of the microscope chamber. The typical electron beam current during the imaging at 300 kV is $10^5$ electrons/nm²s. Because of the low drift of our MEMS heating holder, we were able to use a long exposure time (typically six seconds) to make single-shot images on a 2K by 2K Gatan CCD camera. Image sequences from the movies are performed with a typical interval time of 12 seconds. Then these image sequences are manually aligned and converted into movies via software Image J (Rasband, W.S., *ImageJ*, U. S. National Institutes of Health, Bethesda, Maryland, USA, http://rsb.info.nih.gov/ij/).

**Contamination**

The rate of contamination is measured by putting a 20 nm diameter electron beam of the transmission electron microscope operated at 300 kV on the graphene sample and subsequent monitoring of the increase of amorphous carbon deposition in time. When this contamination test is done at room temperature, a fast carbon deposition can be observed (Figure S2). Continuation of this electron beam irradiation leads to a continuous growth of the contamination. This contamination phenomenon is due to the continuous surface diffusion of hydrocarbons to the electron beam area, where those molecules are cracked to carbon. If the sample is pre-heated at 400°C for several minutes, however, the fast continuous growth of contamination is no longer observed at RT, although still some amorphous material is observed in the electron beam area. After preheating at 400°C but over overnight storage of the specimen at room temperature in the TEM, the build-up of contamination is again much faster and continuous. Pore drilling at room temperature is very slow as can be seen from Fig. S2 - point 3, where hardly any difference with its surrounding can be seen. Pore drilling it is faster at 500°C than at 200° C (see points 4 and 5 in Fig. S2).

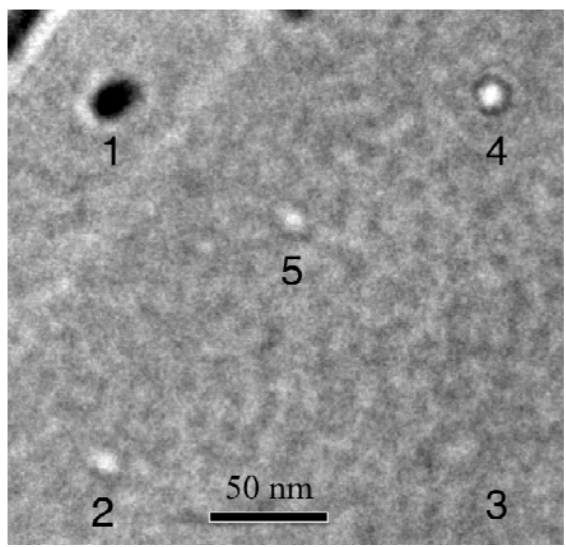

**Figure S2.** Depiction of contamination. Low resolution image (taken at -40 µm underfocus, scale bar 50 nm) showing the effect of a focused electron beam irradiation for 2 minutes at the following conditions

1) Room temperature (RT), 2) 400 °C, 3) RT, 4) 400 °C, and 5) 200 °C for 2 minutes each. Contamination is deposited in 1, holes are created in 2, 4, and 5, and no effect is visible in 3. The image shown here was taken after the whole cycle was performed, and is the same for each point after the experiment at this point was done. Since the C contamination at 1 is still visible after preheating, it is obvious that this contamination can not be removed by specimen heating.

**Fourier transforms of the images of Figure 1 of the main text**

The Fourier transform analyses were made using Image J and its "Radial Profile Plot" function (P. Baggethun, http://rsbweb.nih.gov/ij/plugins/radial-profile.html, 2002-2009).

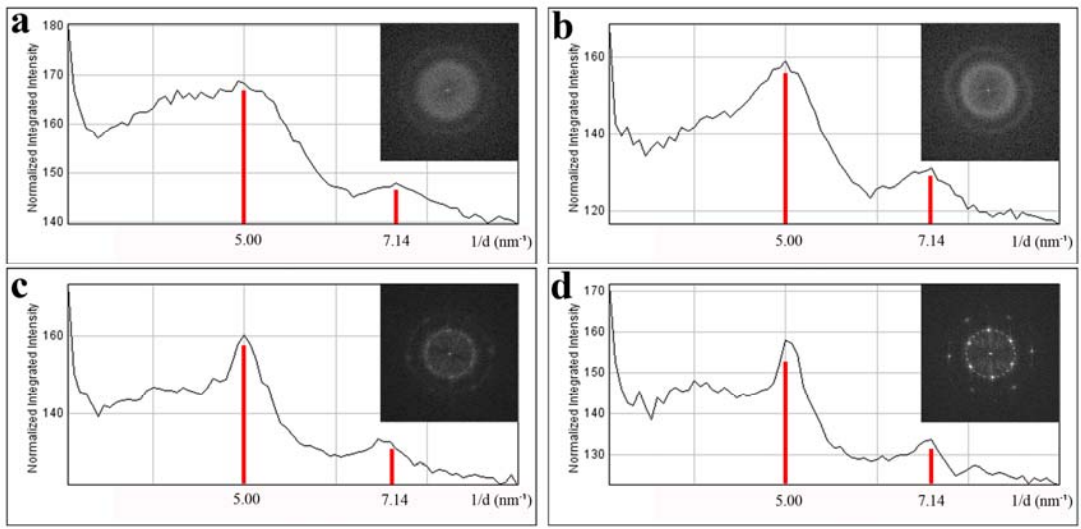

**Figure S3.** Rotational averages of the Fourier Transforms (FTs) of the four images in Fig. 1. The FTs are given as insets. (a) Room temperature (RT), (b) 200 °C, (c) 500 °C, and (d) 700 °C. The plot of RT shows an unpronounced peak at 2 Å, indicating the poor absence of a long- or short-ranged ordered graphene-like structure. The plots of 200, 500 and 700°C show an increasingly sharp peak at 2Å indicating an increasing long-range order. The FT's of 500 and 700 °C show also 100 (5 nm$^{-1}$) and 110 (7.14 nm$^{-1}$) reflections. No ring through these reflections is observed in the FT of 700°C, indicating that the graphene is single-crystalline.

**Recrystallisation at T>500 °C after amorphisation at room temperature**

An amorphous area created by e-beam exposure at room temperature can be transformed into (poly) crystalline graphene by heating to T>500 °C. A hole was made at room temperature as shown in Fig. S4a. The graphene around the hole becomes amorphous due to electron beam irradiation, which can be seen from the Fourier Transform (FT) image (Fig. S4a inset). After increasing the temperature from room temperature to 500 °C, the amorphous area around the hole is polycrystalline as shown in Fig. S4b. The FT rings (Fig. S4b inset) illustrate the polycrystalline graphene-like structure at 500 °C.

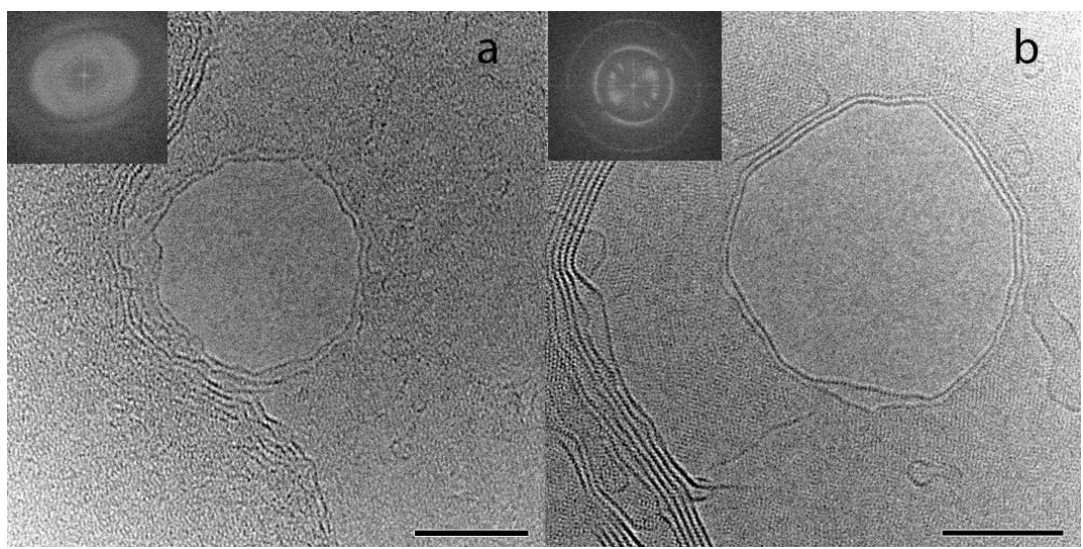

**Figure S4.** Recrystalisation experiments. (a) Hole made at room temperature. (b) The same area heated to 500 °C. Insets show the FTs. Scale bars are 5 nm.

**Straight cylindrical nanopores**

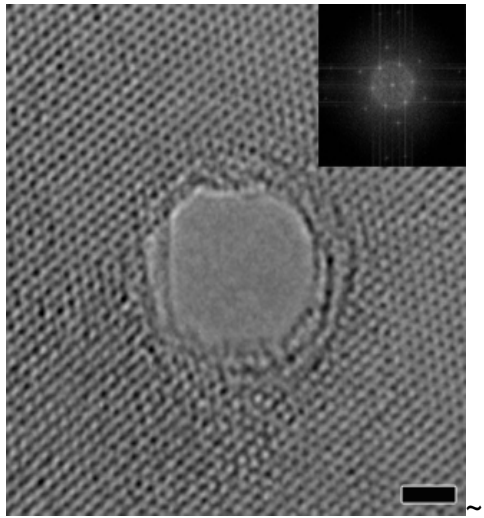

**Figure S5.** Straight cylindrical nanopores made at 600 °C in few-layer graphene. Scale bar, 1 nm. The FT is given as the inset.

**Carbon nanotubes**

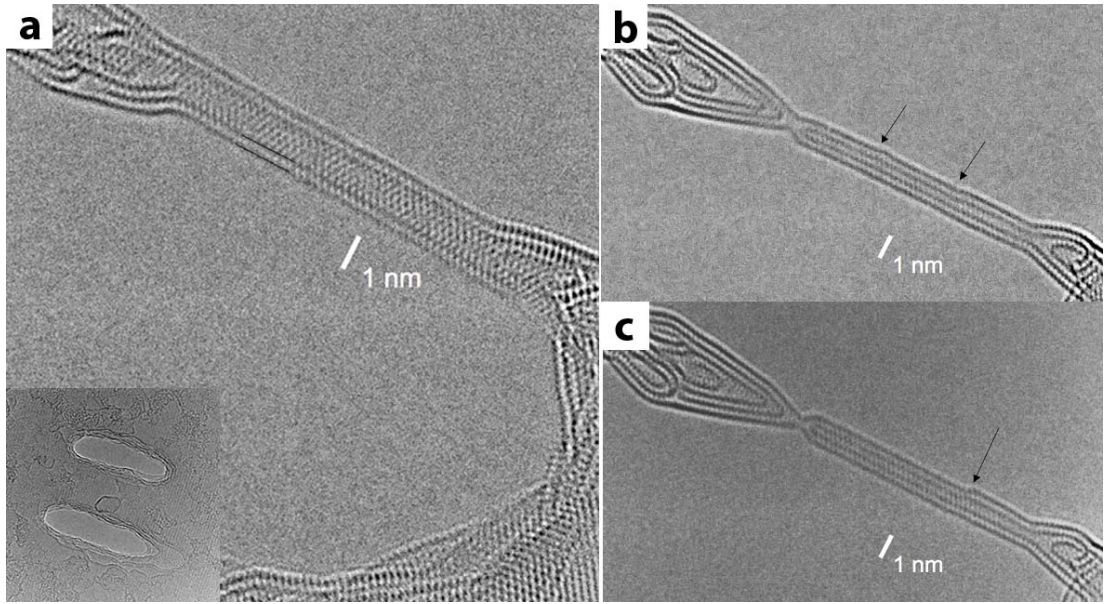

**Figure S6.** Carbon nanotubes made from few-layer graphene at 700 °C. (a) a carbon nanotube made using elongated electron beams with a shape similar to the holes made (inset). (b)-(c) show two frames from a movie (See Movie S3) in which the reduction in width of the CNT can be followed, finally leading a breaking of the CNT. In (a) the tube is about 1.2 nm wide. In (b) two steps in the width of the tube can be observed (indicated by arrows). The inner tube varies from 6 rings to 7 rings and 8 rings with the steps at the arrows. In (c), which is taken 5 seconds later, only a single step (indicated by arrow) in the nanotube width can be observed.

**Polycrystalline graphene nanoribbon**

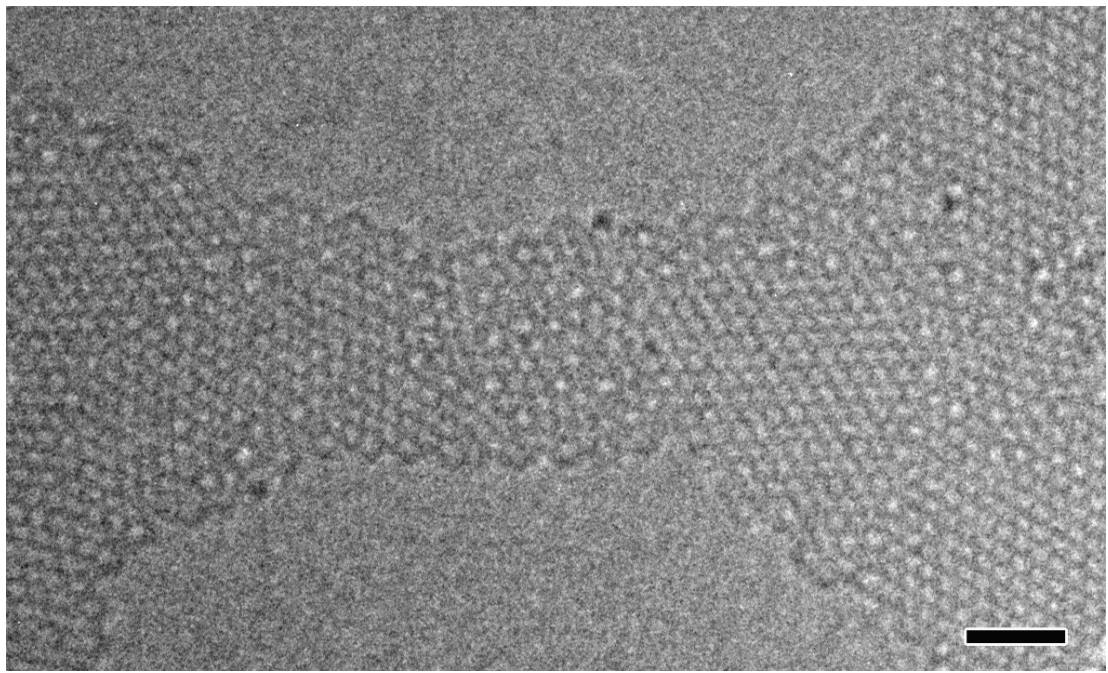

**Figure S7.** A polycrystalline graphene nanoribbon made at 500 °C. Ad-atoms can be seen on the edge and surface of the nanoribbon as dark dots. See Movie S5. Scale bar, 1 nm.

**Prospective: More efficient sculpting of graphene by better hardware**

The e-beam sculpting of graphene can be further optimized by even better hardware. The use of imaging cameras with high detection efficiency like the CMOS cameras, that now become available with allow easier imaging of the sculpted graphene without additional modifications by the electron beam. Note that due to the low contrast of the graphene, one needs to use a relatively high electron dose. The use of a high brightness gun and a probe corrector allow for a smaller high intensity beam and thus a higher precision in the sculpting. Further developments in low drift heating holders will allow longer exposure times and thus the use of less electrons per second.

**Movies**

Movie S1: a polycrystalline monolayer at 500 °C

Movie S2: self-repair at 600 °C

Movie S3: carbon nanotube formation at 600 °C

Movie S4: flat tube formation at 600 °C

Movie S5: polycrystalline bridge formation at 500 °C

Movie S6: Armchair nanoribbon at 700 °C

Movie S7: single carbon chain formation at 600 °C